\documentstyle[aps,prb,twocolumn,epsfig]{revtex}

\begin{document}
\draft \wideabs{
\title{Study of gossamer superconductivity and
antiferromagnetism in the $t$-$J$-$U$ model}
\author{Feng Yuan$^{1,2}, $ Qingshan Yuan$^{1,3}$, C.
S. Ting$^{1}$, and T. K. Lee$^{4}$}
\address{$^1$ Texas Center for Superconductivity and
Advanced Materials and Department of Physics,
University of Houston, Houston, TX 77204\\
$^2$ Department of Physics, Qingdao University, Qingdao 266071,
China\\
$^3$ Pohl Institute of Solid State Physics, Tongji University,
Shanghai 200092, China\\
$^4$ Institute of Physics, Academia Sinica, Nankang, Taipei,
Taiwan 11529}

\date{\today}
\maketitle

\begin{abstract}
The d-wave superconductivity (dSC) and antiferromagnetism are
analytically studied in a renormalized mean field theory for a two
dimensional $t$-$J$ model plus an on-site repulsive Hubbard
interaction $U$. The purpose of introducing the $U$ term is to
partially impose the no double occupancy constraint by employing
the Gutzwiller approximation. The phase diagrams as functions of
doping $\delta$ and $U$ are studied. Using the standard value of
$t/J=3.0$ and in the large $U$ limit, we show that the
antiferromagnetic (AF) order emerges and coexists with the dSC in
the underdoped region below the doping $\delta\sim0.1$. The dSC
order parameter increases from zero as the doping increases and
reaches a maximum near the optimal doping $\delta\sim0.15$.  In
the small $U$ limit, only the dSC order survives while the AF
order disappears. As $U$ increased to a critical value, the AF
order shows up and coexists with the dSC in the underdoped regime.
At half filing, the system is in the dSC state for small $U$ and
becomes an AF insulator for large $U$. Within the present mean
field approach, We show that the ground state energy of the
coexistent state is always lower than that of the pure dSC state.
\end{abstract}
 }

\section{Introduction}

In spite of tremendous theoretical and experimental efforts
dedicated to the studies of the anomalous properties of high
$T_{c}$ superconductors (HTS), a full understanding of these
materials is still far from the final stage. As a basic point, it
is known that much of the physics should come from the competition
between the d-wave superconductivity (dSC) and antiferromagnetism.
Experimentally, it is generally suggested that the ground state
evolves from the antiferromagnetic (AF) state to that of the dSC
order as the carrier density increases\cite{Anderson97}. However,
since the early days of HTS, there also have been persistent
reports of the coexistence of the dSC and AF
orders\cite{Weidinger,Kiefl,Suzuki,Kimura,Sidis,Hodges,Mook}in
various cuprate samples. Especially in the recent neutron
scattering experiments, the commensurate AF order has been
observed in the underdoped superconducting
YBa$_{2}$Cu$_{3}$O$_{6.5}$, providing the unambiguous evidence for
an unusual spin density wave state coexisting with
superconductivity (dSC)\cite{Sidis}. Therefore it is necessary to
develop a microscopic theory in which both the antiferromagnetism
and the dSC are treated equally in order to understand the ground
state property of the cuprate superconductors.

Theoretically, it has  been widely accepted that the essential
physics of cuprates can be effectively described by the two
dimensional Hubbard model or its equivalent $t$-$J$ model in the
large $U$ limit\cite{Anderson87,Zhang88}. Using the variational
Monte Carlo (VMC) method, several groups proposed wave functions
with coexisting  AF and dSC orders and found that the coexisting
state has a lower energy than either the pure dSC order or the
pure AF state in the underdoped
regime\cite{Chen90,Giamarchi,Himeda,Shih}. Although the slave
particle mean field theory for the $t$-$J$ model was originally
introduced to investigate the formation of the RVB state or the
superconducting order
\cite{Anderson87,Baskaran,Kotliar,Zou,Suzumura}, it also has been
applied to study  the coexistence of the dSC and AF orders in this
system\cite{Inaba,Yamase}. Stimulated by  the idea of the
``gossamer superconductors'' proposed by Laughlin\cite{Laughlin},
Zhang and co-workers\cite{Zhang03} employed  the $t$-$J$-$U$ model
with the Gutzwiller projected wave function\cite{Gutzwiller}to
investigate the superconducting order parameter and the electron
pairing gap (or the RVB order parameter). There\cite{Zhang03} the
on-site Coulomb interaction $U$ is introduced to partially impose
the no double occupancy constraint for the strongly correlated
electron systems. In the large $U$ limit, their
result\cite{Zhang03} is consistent with that of Kotliar and
Liu\cite{Kotliar}using the slave boson mean field approach for the
$t$-$J$ model.

Following Ref. [22], we report a further investigation of the same
model by taking the AF order explicitly into consideration. Within
the Gutzwiller renormalized mean field theory, we find that for
large Coulomb repulsion $U$, there is a coexistence between AF and
dSC orders below the doping level $\delta\sim0.1$. The coexisting
state always has a lower energy than  that of the pure dSC state.
The dSC order parameter increases from zero as the doping
increases in the underdoped region and then reaches a maximum near
the optimal doping $\delta\sim0.15$, after that it decreases to
zero at $\delta\sim 0.35$ with increasing doping. When the
magnitude of $U$ is reduced, the AF order parameter decreases very
quickly with increasing doping, and the coexistent region is
squeezed toward low doping regime until it disappears for $U<
5.3t$, where the "gossamer superconductivity" is found even at
half filling.

The paper is organized as follows. In Sec. II, we outline the
theoretical framework. The $t$-$J$-$U$ model is introduced
and the Gutzwiller variational approach is formulated.
A renormalized Hamiltonian is obtained and further studied
within the mean field theory.
In Sec. III, our numerical results are displayed
and compared with those from other theories and experiments.
In Sec. IV, a summary of the paper will be given.

\section{Theoretical framework}

We start from the $t$-$J$-$U$ model on a square lattice\cite{Zhang03},
\begin{eqnarray}
H&=&H_{t}+H_{s}+H_{U},
\end{eqnarray}
with
\begin{eqnarray}
H_{t}&=&-t\sum_{i\hat{\eta}\sigma}(C^{\dagger}_{i\sigma}
C_{i+\hat{\eta}\sigma}+{\rm h.c.}),\nonumber\\
H_{s}&=&J\sum_{i\hat{\eta}} {\bf
S}_{i}\cdot {\bf S}_{i+\hat{\eta}},\nonumber\\
H_{U}&=&U\sum_{i}\hat{n}_{i\uparrow} \hat{n}_{i\downarrow},
\end{eqnarray}
where  $\hat{\eta}=\hat{x}$ and $\hat{y}$,
$C^{\dag}_{i\sigma}(C_{i\sigma})$ is the electron creation
(annihilation) operator, ${\bf
S}_{i}=\sum_{\sigma\sigma'}C^{\dagger}_{i\sigma}
{\vec\sigma}_{\sigma\sigma'}C_{i\sigma'}/2$ is the spin operator
with ${\vec \sigma}=(\sigma_{x},\sigma_{y},\sigma_{z})$ as Pauli
matrices, $\hat{n}_{i\sigma}=C^{\dagger}_{i\sigma}C_{i\sigma}$,
$U$ is the on-site Coulomb repulsion, $t$ is the hopping
parameter, and $J$ is the exchange coupling constant. In the
Hamiltonian (1), the $U$ term is introduced to partially impose
the no double occupancy constraint. In the limit
$U\rightarrow\infty$, the model is reduced to the $t$-$J$ model.

To study the Hamiltonian (1) with the Gutzwiller variational
approach, we take the trial wave function $|\psi\rangle $ as

\begin{eqnarray}
|\psi\rangle=P_{G}|\psi_{0}(\Delta_{d},\Delta_{af},\mu)\rangle,
\end{eqnarray}
where $P_{G}$ is the Gutzwiller projection operator and it is
defined as
\begin{eqnarray}
P_{G}=\Pi_{i}[1-(1-g)\hat{n}_{i\uparrow}\hat{n}_{i\downarrow}],
\end{eqnarray}
here $g$ is a variational parameter which takes the value between
0 and 1. The choice $g=0$ corresponds to the situation with no
doubly occupied sites($U\rightarrow\infty$), while $g=1$
corresponds to the uncorrelated state($U=0$).
$|\psi_{0}(\Delta_{d},\Delta_{af},\mu)\rangle $ is a Hartree-Fock
type wave function, where $\Delta_{d},\Delta_{af},\mu$ are the
parameters representing dSC, antiferromagnetism and chemical
potential, respectively. The nature of $|\psi_{0}\rangle$ depends
on the expected long range behavior. Since it is the purpose of
this paper to study the interplay between antiferromagnetism and
dSC, we will adopt the wave function which includes both the dSC
and antiferromagnetism in a unique variational
space\cite{Giamarchi,Himeda}.

With help of the trial wave function (3), the variational energy
$E_{var}=\langle H \rangle$ is given by
\begin{eqnarray}
E_{var}=\frac{\langle\psi\mid H
\mid\psi\rangle}{\langle\psi\mid\psi\rangle}=NUd+\langle
H_{t}\rangle+\langle H_{s}\rangle,
\end{eqnarray}
where
\begin{eqnarray}
\langle H_{t}\rangle=\frac{\langle\psi\mid H_{t}
\mid\psi\rangle}{\langle\psi\mid\psi\rangle},\nonumber\\
\langle H_{s}\rangle=\frac{\langle\psi\mid H_{s}
\mid\psi\rangle}{\langle\psi\mid\psi\rangle},
\end{eqnarray}
$N$ is the total number of the lattice sites and $d=\langle
n_{i\uparrow}n_{i\downarrow}\rangle$ is the average double
occupation number. Obviously, the double occupancy can be
modulated by $U$.

In the calculation of the variational energy, we adopt the
Gutzwiller projection method which was formulated originally for
the Hubbard Hamiltonian. A clear and simple
explanation\cite{Ogawa} was given by Ogawa {\it et al.} and by
Vollhardt. In their scheme, the spatial correlations are
neglected, and the effect of the projection operator is taken into
account by the classical statistical weight factors. In this way,
the hopping average and the spin-spin correlation in the state
$\mid\psi \rangle $ are related to those in the state
$\mid\psi_{0}\rangle$ through the following relations
\begin{eqnarray}
\frac{\langle\psi\mid C^{\dag}_{i\sigma}C_{j\sigma}
\mid\psi\rangle}{\langle\psi\mid\psi\rangle}=g_{t}\langle\psi_{0}\mid
C^{\dag}_{i\sigma}C_{j\sigma} \mid\psi_{0}\rangle,
\nonumber\\
\frac{\langle\psi\mid S_{i}\cdot S_{j}
\mid\psi\rangle}{\langle\psi\mid\psi\rangle}=g_{s}\langle\psi_{0}\mid
S_{i}\cdot S_{j} \mid\psi_{0}\rangle.
\end{eqnarray}
In the thermodynamic limit, one has \cite{Ogawa}
\begin{eqnarray}
g^{2}=\frac{d(1-n+d)}{(1-r)(1-w)wr}\frac{(n-2wr)^{2}}{(n-2d)^{2}},
\end{eqnarray}
and the renormalization factors can be derived to have the
following expressions,
\begin{eqnarray}
g_{t}=\frac{n-2d}{n-2rw}\left[\sqrt{\frac{(1-w)(1-n+d)}{1-r}}
+\sqrt{\frac{w}{r}d}\right]\nonumber \\
\times
\left[\sqrt{\frac{(1-r)(1-n+d)}{1-w}}+\sqrt{\frac{r}{w}d}\right],
\end{eqnarray}
\begin{eqnarray}
g_{s}=(\frac{n-2d}{n-2wr})^{2}.
\end{eqnarray}
Here $n$ is the average electron number per site. In order to
consider the AF order, the square lattice is divided into two
sublattices $A$ and $B$. For sublattice $A$ we assume $\langle
\hat{n}_{i\uparrow}\rangle \equiv r =\frac{n}{2}+m$ and $ \langle
\hat{n}_{i\downarrow}\rangle \equiv w =\frac{n}{2}-m$, i.e., a net
magnetization $+m$ at each site. For sublattice $B$ the electron
occupation numbers $r$ and $w$ are exchanged, meaning the
magnetization $-m$ at each site. Here $m$ represents the AF order
parameter in the state $|\psi_{0}\rangle$. These renormalization
factors, $g_{t}$ and $g_{s}$, quantitatively describe the
correlation effect of the on-site repulsion. We will further
comment on this point below.

In terms of these renormalization factors, the variational energy
$E_{var}=\langle H \rangle$ is rewritten as
\begin{eqnarray}
E_{var}=\langle H_{eff} \rangle_{0},
\end{eqnarray}
where $H_{eff}$ is the Gutzwiller renormalized Hamiltonian:
\begin{eqnarray}
H_{eff}&=& g_{t}H_{t}+g_{s}H_{s}+H_{U}\nonumber \\
       &=&
-g_{t}t\sum_{i\hat{\eta}\sigma}(C^{\dagger}_{i\sigma}
C_{i+\hat{\eta}\sigma}+{\rm h.c.})\nonumber \\
&+&g_{s}J\sum_{i\hat{\eta}} {\bf S}_{i}\cdot {\bf
S}_{i+\hat{\eta}}+NUd.
\end{eqnarray}

In the mean field approximation the renormalized Hamiltonian (12)
can be rewritten as
\begin{eqnarray}
H_{MF}&=&NUd
+\frac{3}{4}Ng_{s}J(\Delta^{2}+\chi^{2})+2Ng_{s}Jm^{2}\nonumber \\
&+&{\sum_{k\sigma}}'\{(\epsilon_{k}-\mu)C^{\dag}_{k\sigma}C_{k\sigma}
+(\epsilon_{k+Q}-\mu)C^{\dag}_{k+Q\sigma}C_{k+Q\sigma}\nonumber \\
&  &-\sigma\Delta_{af}
(C^{\dag}_{k\sigma}C_{k+Q\sigma}+C^{\dag}_{k+Q\sigma}C_{k\sigma})\}\nonumber
\\
&-&{\sum_{k}}'\Delta_{d}\eta_{k}(C_{-k\downarrow}C_{k\uparrow}-C_{-k+Q\downarrow}C_{k+Q\uparrow}\nonumber
\\
&
&+C^{\dag}_{k\uparrow}C^{\dag}_{-k\downarrow}-C^{\dag}_{k+Q\uparrow}C^{\dag}_{-k+Q\downarrow}),
\label{Hmf}
\end{eqnarray}
where the electron chemical potential $\mu$ has been added,
$Q=(\pi,\pi)$ is the commensurate nesting vector, and the prime on
the summation symbol indicates that $k$ is limited to half of the
Brillouin zone. In the above equation, we have introduced
respectively the electron pairing order parameter, the
hopping average and the staggered magnetization
\begin{eqnarray}
\Delta_{\eta} &= & \langle
C_{i\downarrow}C_{i+\eta\uparrow}-C_{i\uparrow}C_{i+\eta\downarrow}\rangle_{0}\nonumber\\
& = & \Delta\ (-\Delta)\ {\rm when}\ \eta=x\ (y)
\ ,\label{defD}\\
\chi_{\eta}&=&\chi=\langle C^{\dag}_{i\uparrow}C_{i+\eta\uparrow}+
C^{\dag}_{i\downarrow}C_{i+\eta\downarrow}\rangle_{0}\ ,\\
m & = & (-1)^i\langle C^{\dag}_{i\uparrow}C_{i\uparrow}
-C^{\dag}_{i\downarrow}C_{i\downarrow} \rangle_{0}/2\ ,\label{defm}
\end{eqnarray}
with $\gamma_{k}= 2(cosk_{x}+cosk_{y})$, $\eta_{k}=
2(cosk_{x}-cosk_{y})$, $\epsilon_{k}=
-(g_{t}t+\frac{3}{8}g_{s}J\chi)\gamma_{k}$,
$\Delta_{d}=\frac{3}{8}g_{s}J\Delta$, and $\Delta_{af}= 2g_{s}Jm$.
Here the parameter $\Delta_{d}$ is always associated with the
factor $\eta_{k}$ in Eq. (13), which implies that the
superconductivity has a d-wave like symmetry. The mean field
Hamiltonian (\ref{Hmf}) is easily diagonalized, giving rise to
four bands, $\pm E_{1k}$ and $\pm E_{2k}$ with
\begin{eqnarray}
E_{1k}&=& \sqrt{(\xi_{k}-\mu)^{2}+(\Delta_{d}\eta_{k})^{2}}\ ,\nonumber
\\
E_{2k}&=& \sqrt{(-\xi_{k}-\mu)^{2}+(\Delta_{d}\eta_{k})^{2}}\
,\nonumber \\
\xi_{k}&=& \sqrt{\epsilon_{k}^{2}+\Delta_{af}^{2}}\ .
\end{eqnarray}
Here $\Delta_{d}\eta_{k}$ and $\Delta_{af}$ can be regarded
respectively as the energy gap associated with the dSC and the AF
order parameter. The ground state energy is given by
\begin{eqnarray}
E_{var}/N=
Ud&-&\mu\delta-\frac{1}{N}{\sum_{k}}'(E_{1k}+E_{2k})\nonumber \\
&+&\frac{3}{4}g_{s}J(\Delta^{2}+\chi^{2})+2g_{s}Jm^{2}\ .
\end{eqnarray}

By minimizing the ground state energy, we can obtain the
self-consistent equations for the quantities $\Delta$ (the
electron pairing order parameter), $\chi$, $m$ (staggered
magnetization), $d$ and the chemical potential $\mu$ as follows

\begin{eqnarray}
\Delta&=& \frac{1}{4N}{\sum_{k}}'\eta^{2}_{k}\Delta_{d}(\frac{1}
{E_{1k}}+\frac{1}{E_{2k}}), \\
\chi&=&
\frac{1}{4N}{\sum_{k}}'\gamma_{k}\frac{\epsilon_{k}}{\xi_{k}}
(-\frac{\xi_{k}-\mu}{E_{1k}}+\frac{-\xi_{k}-\mu}
{E_{2k}}), \\
m&=& \frac{1}{2N}{\sum_{k}}'\frac{\Delta_{af}}{\xi_{k}}
(\frac{\xi_{k}-\mu}{E_{1k}}-\frac{-\xi_{k}-\mu} {E_{2k}})\nonumber\\
&&-\frac{1}{4Ng_{s}J}(\frac{\partial E_{var}}{\partial g_{t}}
\frac{\partial g_{t}}{\partial m}+\frac{\partial E_{var}}{\partial
g_{s}}
\frac{\partial g_{s}}{\partial m}), \\
 0&=&UN+\frac{\partial E_{var}}{\partial g_{t}}
\frac{\partial g_{t}}{\partial d}+\frac{\partial E_{var}}{\partial
g_{s}} \frac{\partial g_{s}}{\partial d},\\
\delta&=& \frac{1}{N}{\sum_{k}}'(\frac{\xi_{k}-\mu}{E_{1k}}
+\frac{-\xi_{k}-\mu}{E_{2k}}).
\end{eqnarray}

For each doping $\delta$, all the parameters $\Delta$, $\chi$,
$m$, $d$ and $\mu$ are determined self-consistently by the Eqs.
(19)-(23).

\section{Results and Discussion}

Now we summarize our results.  Firstly we discuss the average
double occupation number $d$ as a function of $U$. Our calculated
results at the doping $\delta=0.0$ (solid line), $0.05$ (dashed
line) and $0.1$ (dotted line) for the parameter $t/J=3.0$ at the
temperature $T$=0 are shown in Fig. 1. We find that the average
double occupation number $d$ at $\delta=0.0$ decreases linearly as
function of $U$ till $U=9.3t$, where $d$ shows the similar
behavior of discontinuity as reported  in Ref. [22]. But for the
doped cases, our numerically obtained $d$ as functions of $U$ do
not show this discontinuity, and they become flattened and
decrease slowly at large $U$.

The Gutzwiller renormalization factors $g_{t}$ and $g_{s}$ as
functions of doping $\delta$ for the parameters $t/J=3.0$ and
$U=20t$ at $T=0$ are shown in Fig. 2. The dashed lines are the
corresponding results when the AF order is not considered or $m$
is fixed to zero.  As we mentioned in Sec. II, these factors
quantitatively reflect the partially enforced no double occupancy
constraint due to the on-site Coulomb repulsion U. For large $U$,
the effect of the Gutzwiller projector operators is to reduce the
kinetic energy and enhance the spin-spin correlation. We find that
at low doping, the AF order suppresses the magnitude of $g_{s}$
while $g_{t}$ is only slightly affected.
\begin{figure*}[ht]
\begin{center}
\epsfig{file=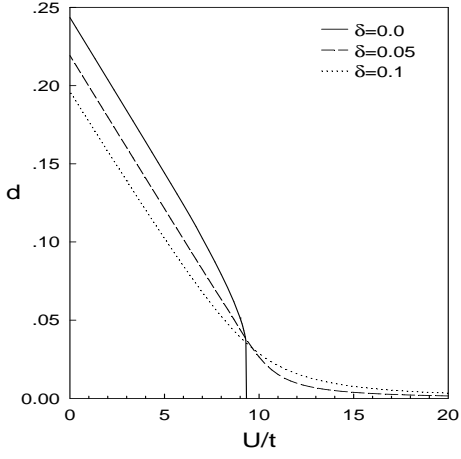,width=6.5cm,height=6.4cm,clip=}
\end{center}
\caption{The average double occupation number $d$ as a function of
$U$ at doping $\delta=0.0$ (solid line), $0.05$ (dashed line), and
$0.1$ (dotted line) for the parameter $t/J=3.0$ at $T$=0. }
\label{Fig1:eps}
\end{figure*}
\begin{figure*}[ht]
\begin{center}
\epsfig{file=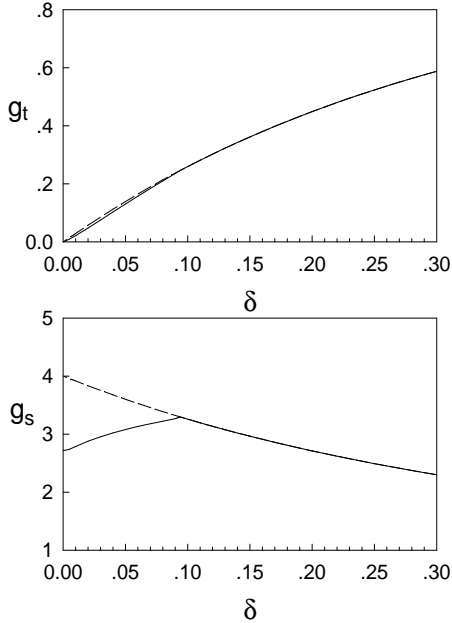,width=6.5cm,height=8.8cm,clip=}
\end{center}
\caption{The Gutzwiller renormalization factors $g_{t}$ and
$g_{s}$ as functions of doping $\delta$ for the parameters $t/J=3.0$
and $U=20t$ at $T=0$ (solid lines). The dashed lines are the corresponding
results when the AF order is not considered, i.e., $m$ is fixed to
zero. } \label{Fig2:eps}
\end{figure*}

In Fig. 3, we plot the self-consistently obtained order parameters
$\Delta$ and $m$ as functions of doping $\delta$ for the
parameters $t/J=3$ and $U=20t$ at $T=0$. The dashed line is the
corresponding $\Delta$ when the staggered magnetization $m$ is set
to zero. It should be noticed that these parameters are the
expectation values under the wave function $|\psi_0\rangle$.  It
is clear that the electron pairing order parameter $\Delta$ is
drastically suppressed at low doping by the AF order. At half
filling, $\Delta$ is reduced to zero and $m$ reaches to its
maximum value. Near $\delta\sim0.1$, the AF order vanishes while
$\Delta$ shows a peak.
\begin{figure*}[ht]
\begin{center}
\epsfig{file=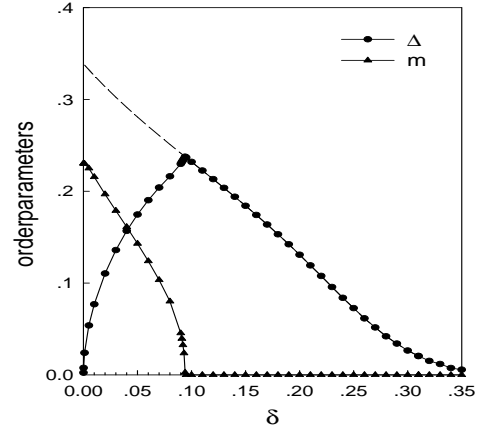,width=6.5cm,height=6.3cm,clip=}
\end{center}
\caption{The self-consistent parameters $\Delta$ and $m$ as
functions of doping $\delta$ for the parameters $t/J=3.0$, $U=20t$
at $T=0$. The dashed line is the $\Delta$ with $m$ is set to
zero.} \label{Fig3:eps}
\end{figure*}
We now discuss the dSC order parameter $\Delta_{SC}$ and AF order
parameter $m_{AF}$ under the wave function $|\psi\rangle$, which
are defined as
\begin{eqnarray}
\Delta_{SC}(\eta) &= & \langle
C_{i\downarrow}C_{i+\eta\uparrow}-C_{i\uparrow}C_{i+\eta\downarrow}\rangle
\nonumber\\
& = & \Delta_{SC}\ (-\Delta_{SC})\ {\rm when}\ \eta=x\ (y)
\ ,\label{defD}\\
m_{AF} & = & (-1)^i\langle C^{\dag}_{i\uparrow}C_{i\uparrow}
-C^{\dag}_{i\downarrow}C_{i\downarrow} \rangle/2\ .\label{defm}
\end{eqnarray}
In the Gutzwiller approximation, these parameters are easily
obtained from $\Delta$ and $m$ with the following renormalization
factors:
\begin{eqnarray}
\Delta_{SC} &= &g_{\Delta}\Delta, \nonumber \\
 m_{AF}&=&g_{m}m.
\end{eqnarray}
Similar to the method of deriving $g_{t}$ and $g_{s}$, we obtain
\begin{eqnarray}
g_{\Delta}=\frac{n-2d}{2(n-2rw)}\left\{\left[\sqrt{\frac{(1-w)(1-n+d)}{1-r}}
+\sqrt{\frac{w}{r}d}\right]^{2} \right.\nonumber \\
\left.
+\left[\sqrt{\frac{(1-r)(1-n+d)}{1-w}}+\sqrt{\frac{r}{w}d}\right]^{2}\right\},
\end{eqnarray}
\begin{eqnarray}
 g_{m}&=&\frac{n-2d}{n-2wr}.
\end{eqnarray}

In Fig. 4 we plot the dSC order parameter $\Delta_{SC}$, AF order
parameter $m_{AF}$ and the electron pairing gap (or the RVB order
parameter\cite{Zhang03}) $\Delta_{d}=\frac{3}{8}g_{s}J\Delta$ as
functions of doping $\delta$ for $t/J=3.0$ and $U=20t$ at $T=0$.
From this phase diagram, we find that the AF and dSC order
parameters coexist for a wide doping range, up to $\delta\sim
0.1$, in the ground state. It can also be seen that the AF order
parameter is a monotonically decreasing function of $\delta$, but
the dSC order parameter shows a non-monotonic dome shape: it
increases from zero as the doping increases in the underdoped
region and then has a maximum near $\delta\sim 0.15$, after which
it decreases to zero at $\delta\sim 0.35$ with increasing doping.
Although the present approach applies only at $T=0$, the
superconducting transition temperature $T_{c}(\delta)$ is expected
to exhibit a similar $\delta$ dependence, and to have a maximum at
the optimal doping $\delta\sim0.15$. It should be noticed that the
electron pairing gap $\Delta_{d}$ is also reduced to zero at half
filling because of the presence of the AF order. This is quite
different from the case in Ref. [22], where the AF order is not
considered, and the electron pairing gap increases as the doping
decreases.

\begin{figure*}[ht]
\begin{center}
\epsfig{file=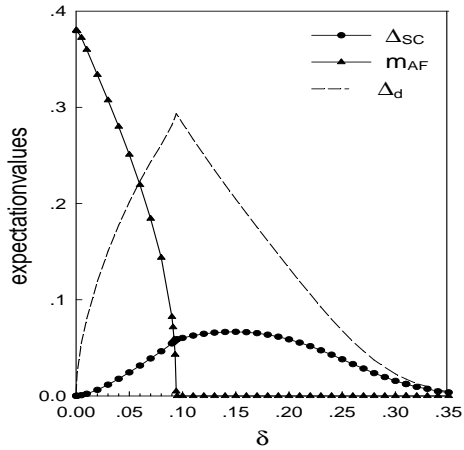,width=6.4cm,height=6.3cm,clip=}
\end{center}
\caption{The dSC order parameter $\Delta_{SC}$, AF order parameter
$m_{AF}$ and the electron pairing gap $\Delta_{d}$ as functions of
doping $\delta$ for $U=20t$ and $t/J=3.0$ at $T=0$.}
\label{Fig:phase}
\end{figure*}

In order to further understand the effect of the Coulomb repulsion
$U$ on the ground state behavior, calculations for several other
values of $U$ are performed. In Fig. 5, we plot the calculated
results for $U=5t$, $7t$, $10t$ and $15t$ with $t/J=3$ and $T=0$.
It is clearly seen that with decreasing $U$, the AF order
decreases very quickly with increasing doping, and the coexistent
region of the AF and dSC orders is squeezed toward lower doping.
Particularly for $U=5t$, the coexistence disappears, and the AF
order is completely suppressed by the prevailing dSC order. To
illustrate more clearly the dependence of the order parameters on
$U$, we present the parameters $\Delta_{SC}$ and $m_{AF}$ as
functions of the Coulomb repulsion $U$ for doping $\delta=0.0$(a),
$\delta=0.05$(b) and $\delta=0.1$(c) at $T=0$ in Fig. 6. At half
filling (see Fig. 6(a)), for small Coulomb repulsion $U < 5.3t$,
only the dSC order persists. As $U$ increases up to $U=5.3t$, the
AF order begins to show up and coexists with the dSC and the
transition appears to be a second order. At $U=7t$, there is a
discontinuity in the slope of $m_{AF}$ and  the dSC order gets
completely suppressed by the AF order at $U>7t$ where our system
becomes an AF insulator. For $U>9.3t$, the double occupancy number
$d$ drops discontinuously to zero. As a result, the magnitude of
$m_{AF}$ jumps from 2.7 to 3.8 and becomes $U$ independent for
large $U$. And with increasing doping (see Fig. 6(b)), the AF
order exists only for  larger $U$ while the dSC order is always in
presence. But for doping $\delta\geq0.1$ (see Fig. 6(c)), the AF
order completely disappears independent of the magnitude of $U$.
\begin{figure*}[ht]
\begin{center}
\epsfig{file=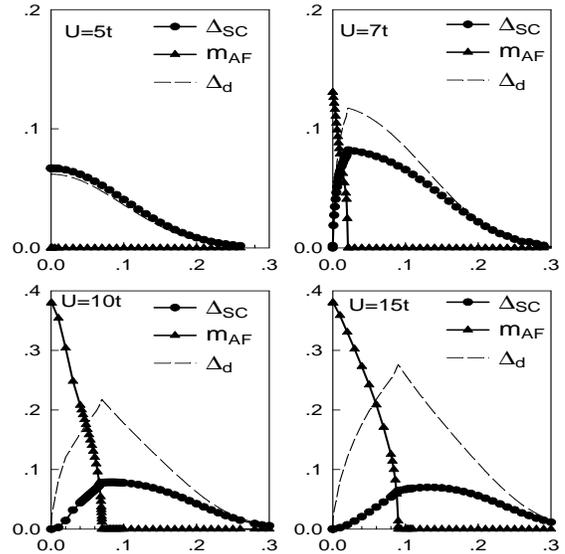,width=8.0cm,height=7.8cm,clip=}
\end{center}
\caption{The dSC order parameter $\Delta_{SC}$, AF order parameter
$m_{AF}$ and the electron pairing gap $\Delta_{d}$ as functions of
doping $\delta$ for different values of $U$ with $t/J=3.0$ and
$T=0$.} \label{Fig:phase2}
\end{figure*}
\begin{figure*}[ht]
\begin{center}
\epsfig{file=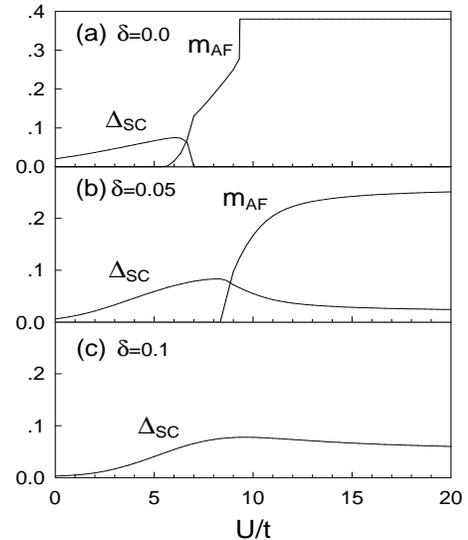,width=6.8cm,height=8.0cm,clip=}
\end{center}
\caption{The dSC and AF order parameters $\Delta_{SC}$ and
$m_{AF}$ as functions of the Coulomb repulsion $U$ for different
dopings with $t/J=3.0$ and $T=0$. } \label{Fig:6}
\end{figure*}

With the help of these self-consistent parameters, let us compare
the ground state energy obtained from Eq.(18) with that of Ref.
[22] in which the contribution from  the AF order was neglected.
In Fig.7, we plot our ground state energy $E_{var}/N$ as a
function of doping $\delta$ using the parameter $t/J=3.0$ for
several different values of $U$ (see the solid lines). The dashed
lines here correspond to the results when the contribution from
the AF order is not included, i.e., $m$ is fixed to zero
\cite{Zhang03}. From Fig.7, we conclude that the ground state
energy with the AF order considered is always lower than that
without it.

\begin{figure*}[ht]
\begin{center}
\epsfig{file=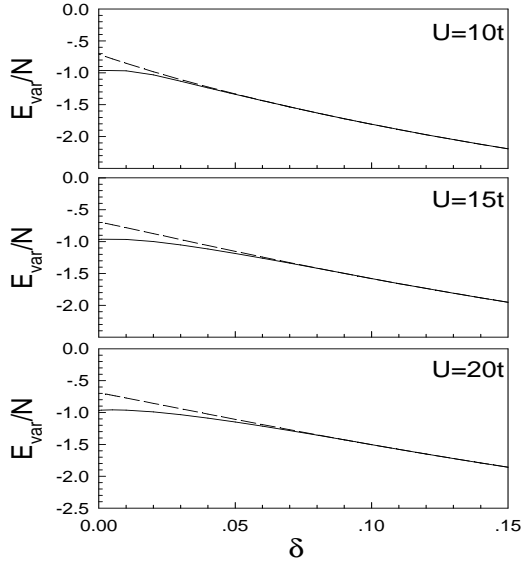,width=7.5cm,height=8.2cm,clip=}
\end{center}
\caption{Doping dependence of the ground state energy for several
different $U$ for the parameter $t/J=3.0$. The dashed lines are
the corresponding results when the AF order is not considered,
i.e., $m$ is fixed to zero. } \label{Fig7:eps}
\end{figure*}

We now discuss the relevance of our calculations to other
theories. Although the $t$-$J$ model, derived from the large $U$
Hubbard model, was originally introduced to study the
superconductivity based on the RVB theory without AF
order\cite{Anderson87,Baskaran,Kotliar,Zou,Suzumura}, the
inclusion of the AF order based on the same approach was done at a
much later stage. In all these studies, the no double occupancy
constraint has been globally enforced. Using the $t$-$J$ or a
similar model and based upon other type of mean field
approximations, there exist several works \cite
{Inaba,Yamase,Inui,Kyung,Lichtenstein,Ogata} investigating the
existence of both AF and dSC orders in the system. While the
double occupancy is globally excluded from the standard $t$-$J$
model, our current $t$-$J$-$U$ model with finite $U$ allows
partial enforcement of the no double occupancy constraint, and to
understand the subtle effect due to the electron-electron
correlation. For the case of small $U$, our results show that only
the dSC order exists in the ground state, which describes the
physics of the "gossamer superconductor". In the limit of infinite
$U$, the $t$-$J$-$U$ model is reduced to the $t$-$J$ model. In
this case our phase diagrams show that the AF and dSC orders
coexist with each other from small $\delta$ up to $\delta
\sim0.1$, and after that the AF order completely disappears. This
feature is in good agreement with the VMC results for the $t$-$J$
model \cite{Giamarchi,Himeda,Shih}. At the same time, we notice
that the coexistence between the AF and dSC orders persists up to
optimal doping $\delta\sim0.15$ in the slave-boson scheme\cite
{Inaba,Yamase}. We would mention that the similar large
coexistence can be obtained if we neglect the derivatives of
$g_{t}$ and $g_{s}$ with $m$ in our derivation of the
self-consistent equations, i.e., replace Eq. (21) with the
following one,
\begin{eqnarray}
m= \frac{1}{2N}{\sum_{k}}'\frac{\Delta_{af}}{\xi_{k}}
(\frac{\xi_{k}-\mu}{E_{1k}}-\frac{-\xi_{k}-\mu} {E_{2k}}).
\end{eqnarray}
In this way, we can perform similar calculations as above. In Fig.
8, we present such a phase diagram with $t/J=3.0$ and $U=15t$ at
$T=0$. It can be seen that in this case, the AF and dSC orders
coexist up to doping $\delta\sim0.18$. But it seems that such a
large coexistent region is not favored by the experimental and
simulation results. Moreover, based on this approximation, the
system at half filing would always be an AF insulator, independent
of the magnitude of $U$. This is contrary to what has been
obtained from our current approach based on minimizing the total
energy of our system.

\begin{figure*}[ht]
\begin{center}
\epsfig{file=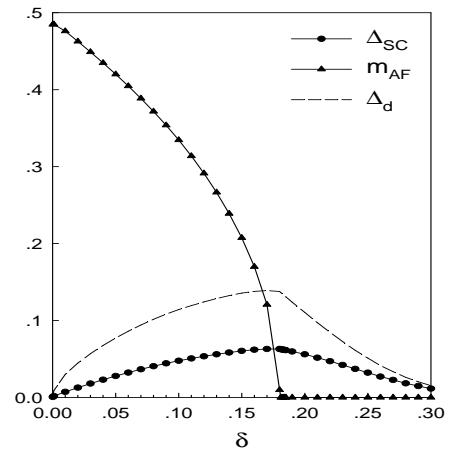,width=6.5cm,height=6.3cm,clip=}
\end{center}
\caption{The dSC order parameter $\Delta_{SC}$, AF order parameter
$m_{AF}$ and the electron pairing gap $\Delta_{d}$ as functions of
doping $\delta$ for $U=15t$ and $t/J=3.0$ at $T=0$. Here the
derivatives of $g_{t}$ and $g_{s}$ with $m$ in the self-consistent
equations are neglected.}\label{Fig:8}
\end{figure*}

So far the experimental evidences for the coexistence of the AF
and dSC orders in cuprate superconductors  seem not conclusive.
For example, the long range AF order observed in the insulating
La$_{2-x}$Sr$_{x}$CuO$_{4}$ is sensitive to
doping\cite{Anderson97}, which disappears rapidly at $x\sim0.03$.
But there also existed several experimental results which appeared
to indicate the coexistence of antiferromagnetism and
superconductivity over a wide doping range in cuprate
superconductors
\cite{Weidinger,Kiefl,Suzuki,Kimura,Sidis,Hodges,Mook}.
Especially, the AF order was claimed to have been observed in
underdoped YBa$_{2}$Cu$_{3}$O$_{6.5}$ and
YBa$_{2}$Cu$_{3}$O$_{6.6}$ superconductors by neutron scattering
experiments from different groups \cite{Sidis,Mook}. It is
apparent that more experiments are needed to confirm the
coexistence of the long range AF order with the dSC state in HTS.

\section{Summary}
In summary, we have studied the coexistence of  the
antiferromagnetism and dSC in a renormalized mean field theory
based on the Gutzwiller approximation for a two dimensional
$t$-$J$-$U$ model. The role of the Hubbard interaction $U$ is to
partially enforce the no double occupancy constraint, and it
provides us with a better understanding of the subtle effect due
to the electron-electron correlation. Our results show that the AF
and dSC orders coexist below the doping $\delta\sim0.1$ at large
$U$ with $t/J=3.0$. And we find that the coexisting state has a
lower ground state energy than that of a pure dSC state. The dSC
order increases from zero as doping increases in the underdoped
regime and reaches a maximum near the optimal doping
$\delta\sim0.15$, after which it decreases to zero at $\delta\sim
0.35$ with increasing doping. With decreasing $U$, the coexistent
region is squeezed toward low doping. There is no coexistence
between AF and dSC orders for small $U$($<5.3t$), where the AF
order is completely suppressed and the "gossamer
superconductivity" is found even at half filling. For the large
$U$, our system at half filling is always an AF insulator in which
both the electron pairing gap and the dSC order parameter are
suppressed to zero. Our result at large $U$ should correspond to
the physical regime. The reason why the existence of the long
range AF order has not been firmly confirmed by experiments in the
underdoped HTS is probably due to  the neglecting of  the AF
fluctuations in the mean field approximation. It is believed that
the effect of the AF fluctuations may break the long range AF
order into short range orders, and this conjecture needs to be
examined more carefully in future theories and experiments on
cuprate superconductors.

\acknowledgments The authors would like to thank  Prof. S. P.
Feng, J. H. Qin, J. Y. Gan, and H. Y. Chen for the helpful
discussions. This work was supported by the Texas Center for
Superconductivity and Advanced Materials at the University of
Houston, and by a grant from the Robert A. Welch Foundation.

\end{document}